\documentclass[twocolumn, english, prl, 
nolongbibliography, superscriptaddress, aps, showpacs, showkeys, amsmath, amssymb, floatfix, nofootinbib]{revtex4-1}

\pdfoutput=1
\usepackage[pdftex]{graphicx}
\usepackage[pdftex]{color}
\usepackage[colorlinks,bookmarks]{hyperref}
\definecolor{linkblue}{rgb}{0,0,0.8}
\definecolor{linkgreen}{rgb}{0,0.5,0}
\hypersetup{linkcolor=linkblue, citecolor=linkgreen, urlcolor=linkblue}

\usepackage[T1]{fontenc}
\usepackage[utf8]{inputenc}
\usepackage{lmodern}
\usepackage{babel}
\usepackage{bm}
\usepackage{esint}
\usepackage{verbatim}
\usepackage[us]{datetime}
\usepackage{booktabs}

\def\d {\mathrm{d}}

\newcommand{\hloc}{H_{0}^\text{local}}

\newcommand{\hlocn}{H_{0, \text{unc}}^\text{local}}
\newcommand{\hcmb}{H_{0}^\text{\tiny CMB}}

\newcommand{\hlocmhcmbdefc}{\Delta H \equiv \left| \hlocn - \hcmb \right|}
\newcommand{\hlocmhcmb}{\Delta H}

\newcommand{\scv}{\sigma_{H_{0}}}

\newcommand{\rhom}{\rho}

\begin{document}

\title{Cosmic variance and the measurement of the local Hubble parameter}

\author{Valerio Marra}
\affiliation{Institut für Theoretische Physik, Ruprecht-Karls-Universität Heidelberg, Philosophenweg
16, 69120 Heidelberg, Germany}

\author{Luca Amendola}
\affiliation{Institut für Theoretische Physik, Ruprecht-Karls-Universität Heidelberg, Philosophenweg
16, 69120 Heidelberg, Germany}

\author{Ignacy Sawicki}
\affiliation{Institut für Theoretische Physik, Ruprecht-Karls-Universität Heidelberg, Philosophenweg
16, 69120 Heidelberg, Germany}

\author{Wessel Valkenburg}
\affiliation{Instituut-Lorentz for Theoretical Physics, Universiteit Leiden, Postbus 9506, 2333 CA Leiden, The Netherlands}
\affiliation{Institut für Theoretische Physik, Ruprecht-Karls-Universität Heidelberg, Philosophenweg
16, 69120 Heidelberg, Germany}

\begin{abstract}
There is an approximately 9\% discrepancy, corresponding to 2.4$\sigma$, between two independent constraints on the
expansion rate of the universe: one indirectly arising from the cosmic
microwave background and baryon acoustic oscillations, and one 
more directly obtained from local measurements of the relation between redshifts
and distances to sources.
We argue that by taking into account the local gravitational potential at the
position of the observer this tension -- strengthened by the recent Planck results -- is \emph{partially}
relieved and the concordance of the standard model of cosmology increased.
We estimate that measurements of the local Hubble constant are
subject to a cosmic variance of about 2.4\% (limiting the local sample to redshifts $z>0.010$) or 1.3\% (limiting it to $z>0.023$), a more significant correction than that taken into account already.
Nonetheless, we show that one would need a very rare fluctuation to fully explain the offset in the Hubble rates. If this tension is further strengthened, a cosmology beyond the standard model may prove necessary.
\end{abstract}


\maketitle

\noindent\paragraph{\bf\emph{Introduction}} 

We can only observe the universe from our own position, which is -- in terms of
cosmological scales -- fixed and lying in a gravitational potential the value of
which possibly cannot be probed~\cite{Valkenburg:2012td}. If the observer could
move around in the universe, they would measure the variation of local
parameters, a variation caused by observing from locations with different values
of the gravitational potential. However, as we cannot measure this unavoidable variation, there is a cosmic variance on physical parameters
that are potentially sensitive to the local spacetime around the observer. One
such parameter is the local expansion rate. 

In this Letter we discuss how the locally measured expansion
rate is offset from the global average expansion rate of the
universe by the value of the gravitational potential at
the observer. By considering the statistics of the distribution of matter in the
universe, we derive the distribution of the gravitational potential at the
observer, and, consequently, the expected distribution of the offset of the
local expansion rate with respect to the global expansion rate. On one hand this
analysis (partially) relieves the tension between existing local and global
measurements of the expansion rate.
On the other hand, our results suggest that local measurements of the Hubble
parameter are
limited to a minimum systematic error of a few percent, which should be included
in the error budget of such measurements.



\noindent\paragraph{\bf\emph{Constraints on the Hubble constant}} 

The most recent measurement of the local Hubble parameter performed by considering recession velocities of objects around us reports a value of
$\hloc=73.8 \pm 2.4$  km\,s${}^{-1}$\,Mpc${}^{-1}$~\cite{Riess:2011yx},
while the Planck 2013 analysis gives
$\hcmb=67.80 \pm 0.77$ km\,s${}^{-1}$\,Mpc${}^{-1}$~\citep[][Table 5]{Ade:2013lta},
assuming a spatially flat  $\Lambda$CDM model (a homogeneous universe with a 
cosmological constant $\Lambda$ and cold dark matter) and fitting to 
observations of the cosmic microwave background (CMB) and baryon acoustic oscillations (BAO) only.
These two independent measurements give a discrepancy of approximately 9\%, corresponding to 2.4$\sigma$.
It is worth stressing that the recent Planck results strengthened this tension, which is only marginal, at 2.0$\sigma$, when the 9-year WMAP data is used~\cite{Hinshaw:2012fq}.
The 9\% disagreement between the expansion rates could be a statistical
fluke or instead a hint for a neglected systematic error. Here we take the
second point of view.
Local fluctuations of the Hubble parameter are indeed to be expected as a
consequence of the density perturbations abundant in the late non-linear
universe.
In particular, a higher $\hloc$ will be observed if we happen to live inside an
underdensity (see e.g.~\cite{Bonvin:2005ps,Hui:2005nm,Cooray:2006ft,Neill:2007fh,Li:2007ny,Hunt:2008wp,Sinclair:2010sb,Umeh:2010pr,Amendola:2010ub,Valkenburg:2011ty,Wiegand:2011je,Romano:2011mx,Valkenburg:2012ds,Marra:2012pj,BenDayan:2012ct,Kalus:2012zu,BenDayan:2013gc,Valkenburg:2013qwa} for studies of the effect of a neglected inhomogeneity on cosmological parameters).
It is therefore natural to ask if the tension between $\hloc$ and $\hcmb$ can
be relieved if a local underdensity consistent with large-scale structure is taken into account in the analysis.

It is interesting to note that the possibility of living in a local underdense ``Hubble bubble'' has been considered before.
Ref.~\cite{Jha:2006fm} found indeed that the Hubble parameter estimated from supernovae Ia (SNe) within 74$h^{-1}$Mpc is $6.5\%\pm1.8\%$ higher than the Hubble parameter measured from SNe outside this region (see also~\cite{Haugboelle:2006uc,Conley:2007ng}).
The analysis of \cite{Riess:2011yx} considers this issue and tries to correct for it; we will discuss this later.
The topic of a local Hubble bubble dates back to the 90's, see e.g.~\cite{1992AJ....103.1427T,Suto:1994kd,Shi:1995nq,Shi:1997aa,Wang:1997tp,Zehavi:1998gz,Giovanelli:1999xp} for previous work on the cosmic variance of the local Hubble parameter.

\noindent\paragraph{\bf\emph{The Hubble bubble model}} 

To tackle this problem we take the simplest approach, that is, we model the
inhomogeneity by means of the Hubble bubble model, which is the basis of the
so-called spherical ``top-hat'' collapse~\cite{Kolb:1990vq}. The idea is to
carve out of the FLRW background a sphere of matter which is then compressed or
diluted so as to obtain a toy model of the inhomogeneity with a slightly different FLRW
solution.
At the junction of the two metrics, the density is discontinuous and the
description could be improved by means of the spherically symmetric
Lema\^{i}tre-Tolman-Bondi (LTB) solution of Einstein's equation~\cite{Lemaitre:1933gd, Tolman:1934za, Bondi:1947av}.
For our purposes, however, the Hubble bubble model suffices, as we are not interested in
the junction between inhomogeneity and background.

\begin{figure}
\includegraphics[width= .9 \columnwidth]{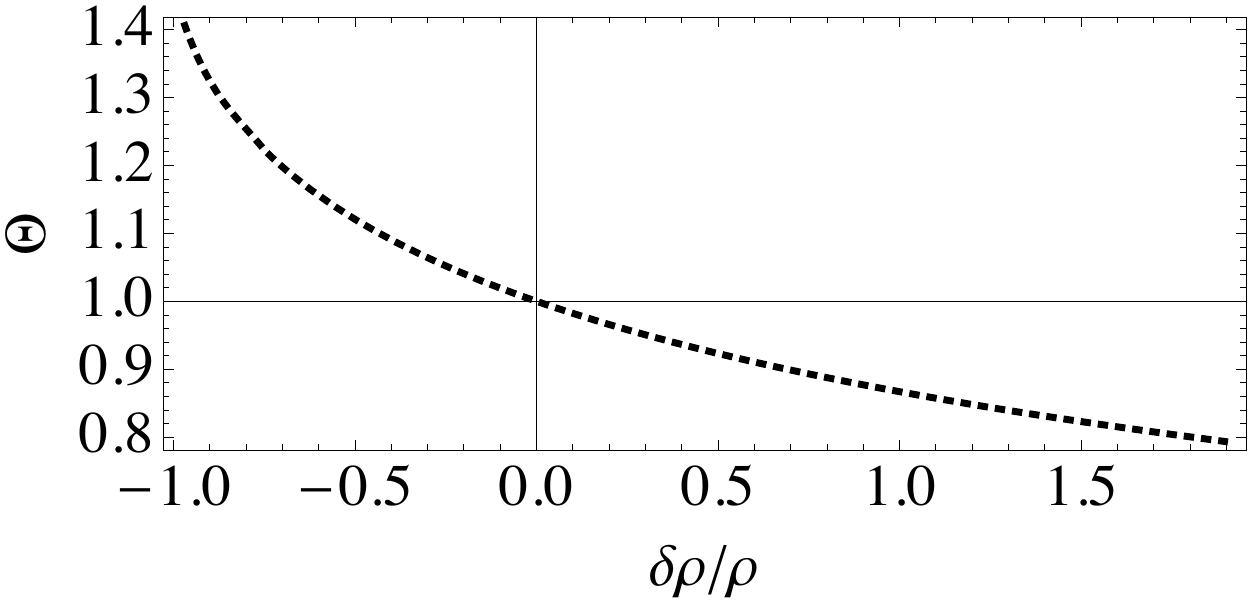}
\caption{Function $\Theta$ which corrects the relation of Eq.~(\ref{relafo}) when the density contrast is not linear. The plot assumes the Planck+BAO best-fit value of $\Omega_{m}=0.3086$, but the dependence of $\Theta$ on cosmological parameters is very weak.}
\label{corre}
\end{figure}

A straightforward prediction of the Hubble bubble model is that an 
adiabatic perturbation in density causes a perturbation in the expansion
rate given by:
\begin{align} \label{relafo}
\frac{\delta H}{H} = -\frac{1}{3} \; \frac{\delta \rhom}{\rhom} \;
f(\Omega_{m}) \; \Theta \! \left(\frac{\delta \rhom}{\rhom},
\Omega_{m} \right) \,,
\end{align}
where all quantities are evaluated at the present time. The function
$f(\Omega_{m})$ is the growth rate and embodies the effect of a
non-negligible cosmological constant\footnote{We assume spatial
flatness so that
$f(\Omega_{m})=-\frac{3}{2}\Omega_{m}+\frac{\frac{15}{16} \Omega_{m}^{1/2}}{_2F_1\left(-\frac{1}{2},\frac{5}{6};\frac{11}{6};1-\Omega_{m}^{-1} \right)-(\frac{3}{8} +\frac{1}{4} \Omega_{m}^{-1}) \, _2F_1\left(\frac{1}{2},\frac{5}{6};\frac{11}{6};1-\Omega_{m}^{-1} \right)}\approx \Omega_{m}^{0.55}$ \cite[see][where also a fit valid for $w\neq -1$ was obtained]{Linder:2005in}, which can be represented in terms of elliptic integrals as in Eq.~(66) of Ref.~\cite{Valkenburg:2011tm}.}. During matter domination one has $f
= 1$, and the standard relation is recovered.
In Fig.~\ref{corre} we show the function $\Theta\left(\tfrac{\delta \rhom}{\rhom},
\Omega_{m} \right)$, which parametrizes the effect of values of $\delta \rhom / \rhom$ approaching the non-linear regime, 
computed by means of the $\Lambda$LTB
model~\cite{Marra:2010pg,Valkenburg:2011tm}.\footnote{For the Planck+BAO best-fit cosmology and the range of contrasts $\delta$ shown in Fig.~\ref{corre}, the function $\Theta$ can be approximated with maximum error of 0.4\% by the fit $\Theta(\delta)=1-0.0882 \, \delta-\frac{0.123 \sin \delta}{1.29+\delta}$.} For linear contrasts, $| \delta \rhom / \rhom| \ll 1$, we have $\Theta \simeq 1$ and Eq.~(\ref{relafo}) becomes a linear relation between perturbations in the density and perturbations in the expansion rate.

The local measurements of the Hubble constant from Ref.~\cite{Riess:2011yx} use standard candles within the redshift range bounded by $ z_{\rm min}=0.010$ (or 0.023) and $ z_{\rm max}=0.1$.
Therefore, we need to know the typical contrast of a perturbation that extends over a redshift in this range.
We take a conservative approach and consider density perturbations stemming from a standard matter power spectrum $P(k)$ with Planck+BAO best-fit parameters.
Consequently, we know that the mean square of the density perturbation in a sphere of 
radius $R$ around any point today -- and so also around us -- is
\begin{align} \label{sigmaR}
\sigma_{R}^{2} \equiv \left( \frac{\delta M}{M} \right)^{2}= \int_0^\infty \frac{k^2 \d k}{2\pi^2} \, P(k)  \left[ \frac{3j_1 (R \, k)}{R \, k} \right]^2 \,,
\end{align}
where $M$ is the mass enclosed by a sphere of radius $R$ and $j_1$ is the spherical Bessel function of the first kind.

\begin{figure}
\includegraphics[width=\columnwidth]{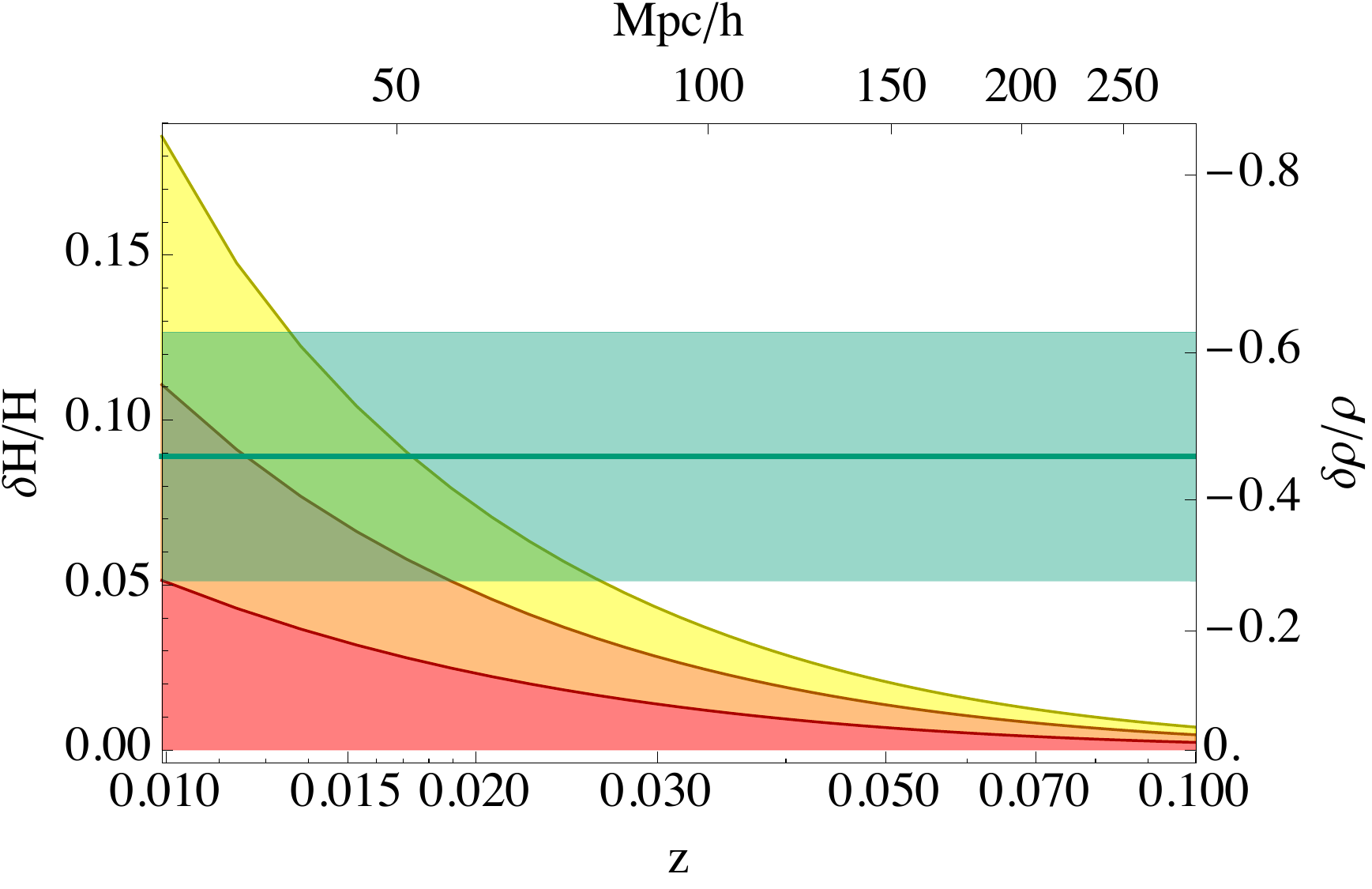}
\caption{The 68\%, 95\% and 99.7\% confidence-level probabilities of gaussian matter fluctuations (right vertical axis) and consequently of the local Hubble parameter (left vertical axis), as a function of co-moving size of the matter fluctuation (top ticks) or, equivalently, redshift (bottom ticks). The relation between $\delta H / H$ and $\delta\rhom/\rhom$ is given by Eq.~\eqref{relafo}.
The range $z_{\rm min}\le z \le z_{\rm max}$ corresponds to the range of observation of~\cite{Riess:2011yx}.
Also shown is the 1-$\sigma$ emerald band relative to the value $\hloc/\hcmb-1$, which shows the 2.4$\sigma$ tension between CMB and local measurements of the Hubble constant.}
\label{figu}
\end{figure}

Next we assume that perturbations in the density field follow a gaussian distribution $p_{\rm gau}$ with the variance given by $\sigma_{R}^{2}$ of Eq.~(\ref{sigmaR}):
\begin{align} \label{gau}
p_{\rm gau}(x) = \frac{1}{\sigma_{R} \sqrt{2 \pi}} \; e^{-\tfrac{x^{2}}{2 \, \sigma_{R}^{2}}} \,,
\end{align}
with $x\equiv\delta\rhom/\rhom$.
In Fig.~\ref{figu} we plot the 68\%, 95\% and 99.7\% confidence-level fluctuations on the local Hubble parameter, as well as the 1-$\sigma$ band relative to the value $\hloc/\hcmb-1$, which shows the 2.4-$\sigma$ tension discussed above.

\begin{figure}
\includegraphics[width=\columnwidth]{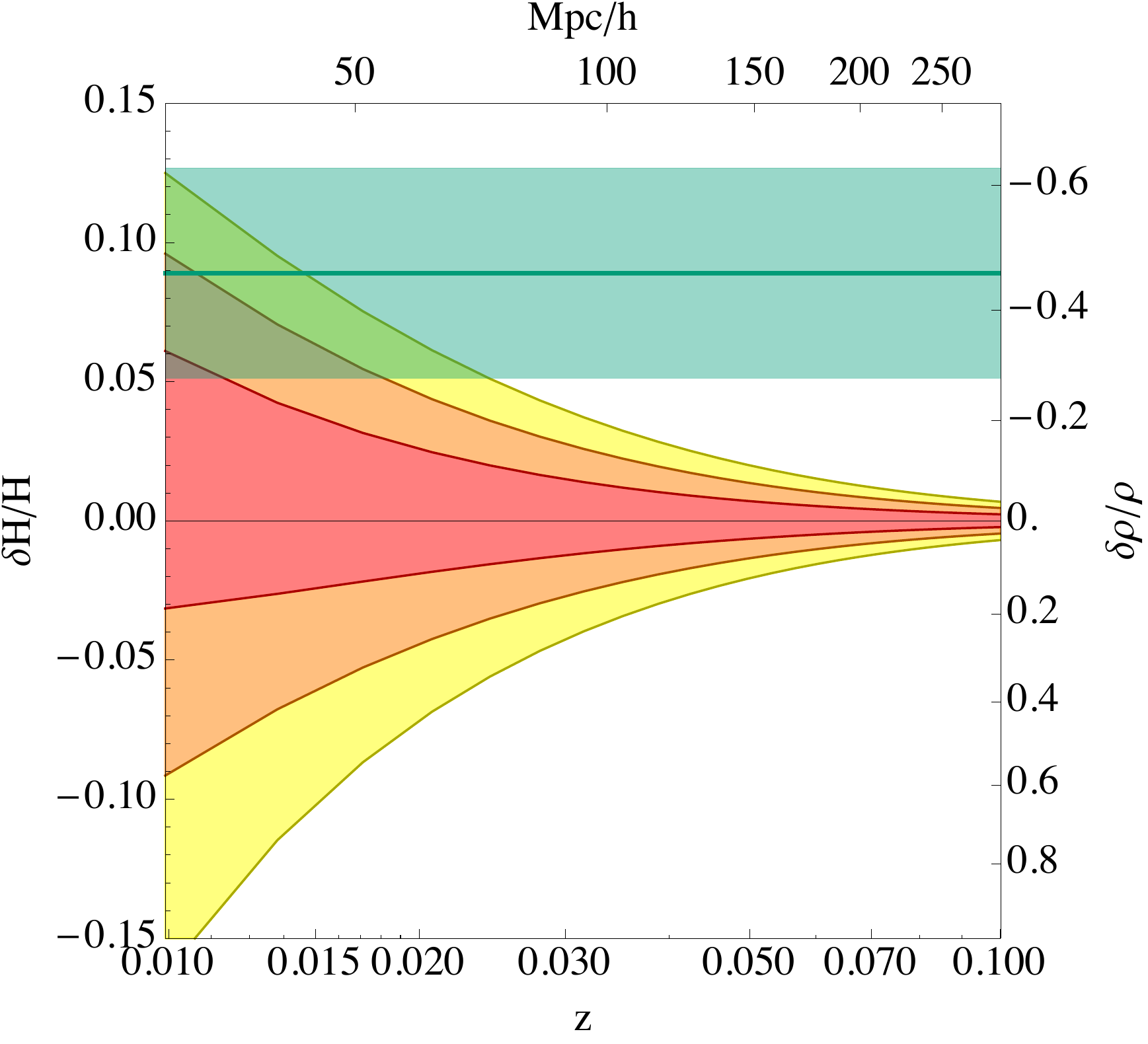}
\caption{The 68\%, 95\% and 99.7\% confidence-level probabilities of log-normally distributed matter fluctuations (right vertical axis) and consequently of the local Hubble parameter (left vertical axis), as a function of co-moving size of the matter fluctuation (top ticks) or, equivalently, redshift (bottom ticks). 
As in Fig.~\ref{figu} we show the 1-$\sigma$ band relative to the value $\hloc/\hcmb-1$.}
\label{figuLN}
\end{figure}

In reality, nonlinear matter fluctuations are better described by a lognormal distribution~\cite{Coles:1991if}:
\begin{align} \label{logn}
p_{\rm logn}(x) = \frac{\exp \left[-\frac{\left(\log \left(\sigma^2_R+1\right)+2 \log (x+1)\right)^2}{8 \log \left(\sigma^2_R+1\right)}\right]}{\sqrt{2 \pi } (x+1) \sqrt{\log \left(\sigma^2_R+1\right)}} \,,
\end{align}
which has zero mean, variance $\sigma^2_R$ and support $(-1,
\infty]$ -- in agreement with the fact that $\delta \rhom/\rhom > -1$.
Moreover, for $\sigma_{R} \rightarrow 0$ it approaches the gaussian distribution
of Eq.~(\ref{gau}).
In Fig.~\ref{figuLN} we show the 68\%, 95\% and 99.7\% confidence level fluctuations of the local Hubble parameter induced by log-normally distributed matter perturbations.
We show separately the case for both over- and under-densities as they are no longer
symmetric when using a skewed distribution such as Eq.~(\ref{logn}). 
Using the log-normal distribution, we see that local voids at a low redshift are actually more likely than they would appear from a gaussian distribution. From here on, we will use the superscripts $+,-$ to refer to the distinct distributions of positive and negative perturbations and their properties, in particular the mean systematic error $\sigma^\pm_{H_0}$. For the symmetric gaussian distribution we of course have $\scv^{+}=\scv^{-}$.

\noindent\paragraph{\bf\emph{Discussion}}

In order to estimate the mean systematic error on local determinations of the Hubble constant we average the 68\% confidence level on $\delta H/H$ over the survey range:
\begin{align} \label{sysh}
\scv^{\pm}
=\left[ \int_{z_{\rm min}}^{z_{\rm max}} \d z \, W_{\rm SN}(z) \, \left( \frac{\delta H^{\pm}}{H^{\phantom{\pm}}} \right)^{2} \right]^{1 \over 2} \,.
\end{align}
In the equation above, the quantity $W_{\rm SN}(z)$ represents the redshift distribution of the SNe used in~\cite{Riess:2011yx}, which is peaked at the lower redshifts.
It is important to stress at this point that we are assuming that the SNe are isotropically distributed over the sky. This implies that we are neglecting the effect of the anisotropic distribution of the sources, which could increase sizably the magnitude of the cosmic variance.
We list in Table \ref{tab:ten} the numerical values of Eq.~(\ref{sysh}) for combinations of cases where either the gaussian distribution of Eq.~\eqref{gau} or the skewed log-normal distribution of Eq.~\eqref{logn} is used.

As $\delta H/H$ is naturally larger at lower redshift, the value of $\scv$ depends strongly on $W_{\rm SN}(z)$ and, in particular, on $z_{\rm min}$ and $z_{\rm max}$.
If one were to extend the upper range $z_{\rm max}$ then the cosmic variance $\scv$ could be reduced at the cost that the uncertainty in the values of the cosmological parameters $\Omega_\text{m},\Omega_\Lambda$, negligible in the current analysis, would begin to play a role.
Alternatively, one could reduce the effect of the cosmic variance by increasing the lower cutoff $z_{\rm min}$.
As discussed earlier, Ref.~\cite{Jha:2006fm} claims that the expansion rate estimated from SNe within 74$h^{-1}$Mpc (corresponding approximately to $z=0.023$) is $6.5\%\pm1.8\%$ larger than the one measured from SNe outside this region.
Consequently, one can alleviate the Hubble bubble effect by adopting $z_{\rm min}=0.023$~\cite{Riess:2011yx}.
In Table \ref{tab:ten}, we also show the values of $\scv$ corresponding to this choice.
The median redshift of the SN redshift distribution is $z_{\rm median}\simeq0.025$ if $z_{\rm min}=0.010$ is used, and $z_{\rm median}\simeq0.033$ if $z_{\rm min}=0.023$ is adopted instead. 
Also, from Figures~\ref{figu} and \ref{figuLN} one can see that this mismatch of 6.5\% can be explained by a local inhomogeneity in agreement with the standard model at about $2\sigma_{R}$.

\begin{table*}[!ht]
\begin{tabular}{ccp{2.4 cm}ccccccp{2cm}p{2cm}}
\toprule
Case && Density Contrast Distribution & $z_\text{min}$ && $\scv^{+}$ & $\scv^{-}$& $ \delta H_{0}^{+} \left (\frac{\rm km/s}{\rm Mpc} \right)$&&  Adding errors linearly   &Adding errors in quadrature\\ \midrule
I &&$p_{\rm gau \phantom{\tiny c \!}}$ of Eq.~(\ref{gau})  & $0.010$ && 2.1\% &2.1\% & 1.58   &&  $  \hlocmhcmb=1.6 \sigma$  & $\hlocmhcmb=2.1 \sigma$  \\
II && $p_{\rm logn}$ of Eq.~(\ref{logn}) & $0.010$ && 2.4\% & 1.7\% & 1.79  &&  $ \hlocmhcmb=1.5 \sigma$  & $\hlocmhcmb=2.1 \sigma$ \\
III && $p_{\rm gau \phantom{\tiny c \!}}$ of Eq.~(\ref{gau})  & $0.023$ && 1.2\% &1.2\%  & 0.90  &&  $  \hlocmhcmb=1.9 \sigma$  & $\hlocmhcmb=2.4 \sigma$  \\
IV && $p_{\rm logn}$ of Eq.~(\ref{logn}) & $0.023$ && 1.3\% & 1.1\% & 0.97  &&  $ \hlocmhcmb=1.8 \sigma$   & $\hlocmhcmb=2.4\sigma$ \\
\bottomrule
\end{tabular}
\caption{Cosmic variance $\scv^{\pm}$ of the local Hubble parameter calculated using Eq.~(\ref{sysh}). $p_\text{gau}$ and $p_\text{logn}$ denote the statistical distribution used to describe the density contrast, $\delta\rhom/\rhom$, gaussian (\ref{gau}) or lognormal (\ref{logn}). $z_\text{min}$ denotes the minimum redshift of the SNe included in the sample. The gaussian distribution has symmetric errors, $\scv^{+}=\scv^{-}$. The quantity $\delta H_{0}^{+}$ gives the absolute error relative to $\scv^{+}$ for $\hloc$. Finally, $\hlocmhcmbdefc = 2.5\sigma$ describes how much the tension between the CMB and local measurement of $H_0$ is reduced when $\scv^{+}$ is included as a systematic error.
The quantity $\hlocn$ is the 0.5\%-larger uncorrected value of the local Hubble constant, see the main text for more details.}
\label{tab:ten}
\end{table*}

It is now natural to ask how much this additional error from the cosmic variance of
our local gravitational potential can relieve the tension of 9\% between the
central values of the two observations discussed at the beginning.
Before proceeding, however, we should point out that Ref.~\cite{Riess:2011yx} besides limiting in most of the analysis the sample to $z_{\rm min}=0.023$, also tries to address the cosmic variance uncertainty by correcting each SN Ia on the Hubble diagram for the expected perturbation of its redshift as determined from the IRAS PSCz density field~\cite{Branchini:1999qj}, in particular by adopting the model B05 of Ref.~\cite{Neill:2007fh}.
The result of this velocity correction causes the final value of $H_0$ to decrease by $0.5\%\pm0.1\%$.
While this approach is in our opinion the right way to proceed so as to deal with the cosmic variance, in light of the tension between $\hcmb$ and $\hloc$ and the uncertainties in the model of Ref.~\cite{Neill:2007fh},\footnote{The analysis of~\cite{Neill:2007fh} depends on the estimate of the bias, assumes a linear relation between velocities and galaxy counts, and is affected by the selection function of the IRAS PSCz density field which drops off at larger scales. Also, the model B05 of~\cite{Neill:2007fh} cannot explain the Hubble bubble detected by~\cite{Jha:2006fm}, which we mentioned at the beginning.} we think it is worth considering the case in which one does not use the results of \cite{Neill:2007fh} and more conservatively estimates the variance stemming from standard inhomogeneities.
We therefore compare the global $\hcmb$ to the 0.5\%-larger uncorrected value of $\hlocn = 74.2 \pm 2.4$  km\,s${}^{-1}$\,Mpc${}^{-1}$.
This slightly increases the tension which is now $\hlocmhcmbdefc = 2.5\sigma$.
As the error from cosmic variance is systematic in nature it should be kept separate from the statistical one.
Just to give a rough estimate, we list in Table~\ref{tab:ten} 
  how much the tension is reduced by adding the errors linearly or in quadrature.
When
using the log-normal distribution we employ the value $\scv^{+}$ as $\hloc > \hcmb$.

\noindent\paragraph{\bf\emph{Conclusions}} 

The simple analysis of this Letter carries two messages.
The first is that local measurements of the Hubble parameter are limited to the minimum systematic error $\delta H_{0}^{+}$ listed in Table \ref{tab:ten}.
These results qualitatively agree with previous estimations of the cosmic variance of the local expansion rate (see e.g.~\cite{Shi:1997aa,Wang:1997tp,Kalus:2012zu}).

\begin{figure}
\includegraphics[width=\columnwidth]{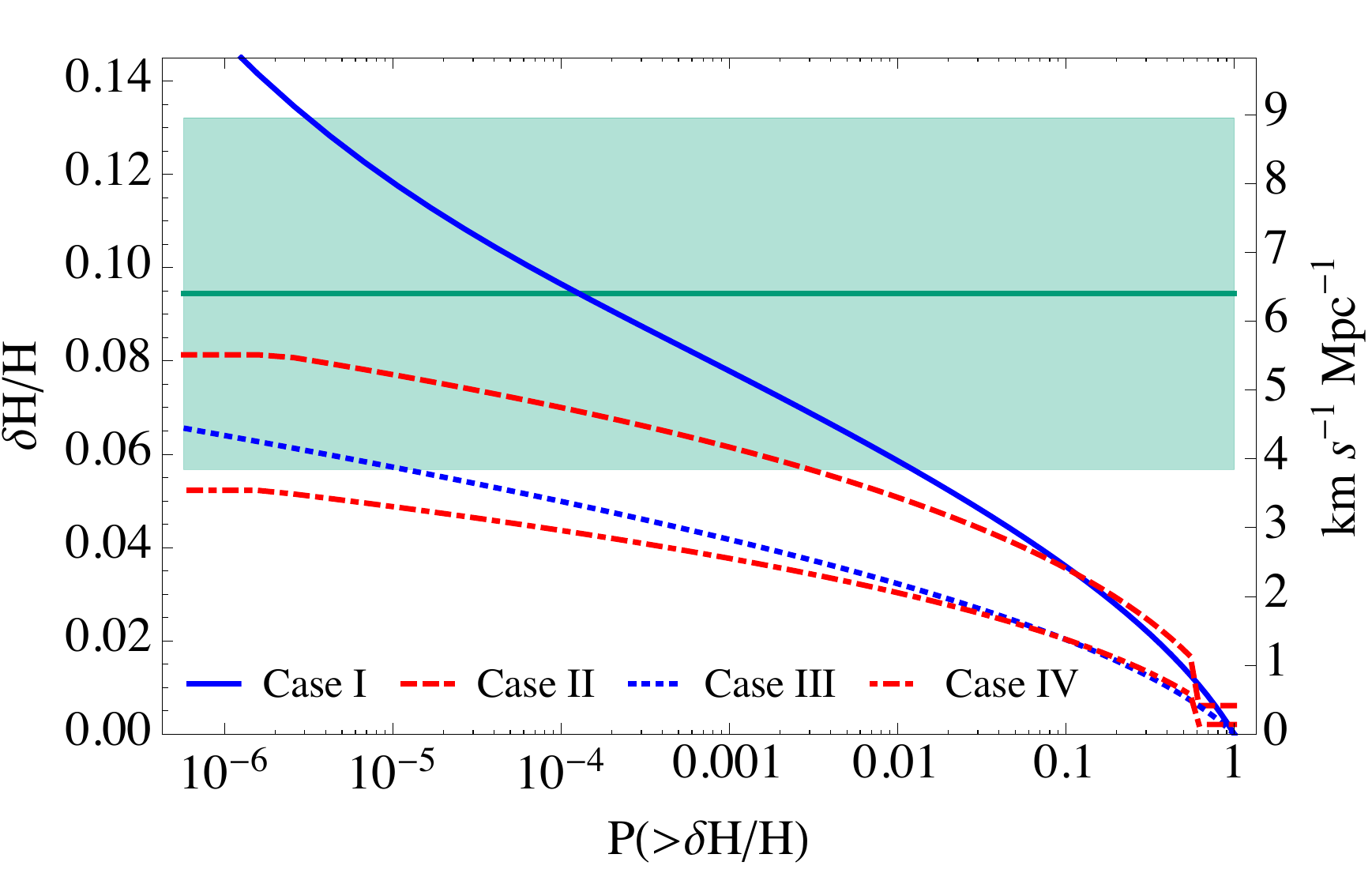}
\caption{Probability of having an inhomogeneity that induces a $\delta H/H$ (left vertical axis) or $\delta H$ (right vertical axis) larger than a given value for the cases listed in the legend and in Table \ref{tab:ten}.
Also shown is the 1-$\sigma$ band relative to the value $\hlocn/\hcmb-1$.}
\label{discoH}
\end{figure}

The second point is that by including the effect of a local inhomogeneity -- in
particular a local underdensity -- the tension between CMB and local measurements
of the Hubble constant is alleviated, even though only partially.
One can quantify the remaining tension by estimating the probability that inhomogeneities stemming from a standard matter power spectrum can explain the 9\% discrepancy. We show in Fig.~\ref{discoH} the result for the four cases discussed in Table \ref{tab:ten}: it is evident that one needs a very rare large-scale structure to explain away the offset in the Hubble rates.
If this tension is further increased,\footnote{Other analyses report higher local Hubble rates, see e.g.~\cite{Riess:2011vh}.} a cosmology beyond the standard model may prove necessary.

Of course, a more thorough analysis is needed in order to precisely quantify the
effect of the local inhomogeneity on measurements of the expansion rate,
possibly by introducing the effect of perturbations of the local gravitational
potential directly in the first steps of the data analysis, as in~\cite{Riess:2011yx}.
Nonetheless, the results of this Letter provide a quick and easy way --
equations (\ref{relafo}) to (\ref{sysh}) -- to estimate  the systematic error
$\scv$, which can be specialized to a given survey by using the corresponding
distribution of standard candles $W_{\rm SN}(z)$.

Finally, in the present era of ``precision'' cosmology it is of crucial importance to fully understand the source of this offset in the Hubble rates, if it is a mere systematic error or new physics.
If one neglects this issue, a fit of a cosmological experiment at large scale combined with local measurements of the Hubble constant biases the extracted cosmological parameters e.g.~the equation of state of dark energy and the effective number of relativistic degrees of freedom. On the other hand, disregarding local measurements on the basis of this disagreement might potentially obscure a hint of cosmology beyond the standard model. This is clearly shown by the analysis of the Planck collaboration, see e.g.~Eqs~(91-93) in~\cite{Ade:2013lta}.

\noindent\paragraph{\bf\emph{Acknowledgements}} 

It is a pleasure to thank
Adam Riess, Dominik Schwarz, David Valls-Gabaud, Licia Verde, James Zibin
for useful comments and discussions.  
The authors acknowledge funding from DFG through the project TRR33 ``The Dark Universe''.

\bibliography{refs}

\end{document}